\documentclass[%
preprint,
superscriptaddress,
amsmath,amssymb,
prx,
]{revtex4-1}
\pdfoutput=1 
\usepackage[english]{babel}
\usepackage{graphicx}
\usepackage{dcolumn}
\usepackage{hyperref}
\usepackage{comment}
\usepackage{xcolor}
\usepackage{bm}
\graphicspath{{Figures/}}
\usepackage{wasysym}
\usepackage{amssymb}
\usepackage{amsmath}
\usepackage{xcolor}
\usepackage{mathtools}

\usepackage[version=4]{mhchem}
\definecolor{greenscatter}{RGB}{53, 183, 121}
\definecolor{bluescatter}{RGB}{49, 104, 142}
\definecolor{purplescatter}{RGB}{68, 57, 131}
\definecolor{greyscatter}{RGB}{178, 178, 178}

\usepackage{acro}
\DeclareAcronym{afq}{short = AFQ, long  = antiferroquadrupolar}
\DeclareAcronym{sc}{short = SC, long  = superconductivity}
\DeclareAcronym{qcp}{short = QCP, long  = quantum critical point}
\DeclareAcronym{cef}{short = CEF, long  = crystal electric field}
\DeclareAcronym{ltl}{short = LTL, long  = Low Temperature Laboratory}
\DeclareAcronym{hfml}{short = HFML, long  = High Field Magnet Laboratory}
\DeclareAcronym{lia}{short = LIA, long  = lock-in amplifier}
\DeclareAcronym{ult}{short = ULT, long  = ultra-low temperature}

\newcommand{\ham}{\ensuremath\hat{\mathcal{H}}}

\begin{document}

\title{Diverse influences of hyperfine interactions on strongly correlated electron states }
\author{Femke Bangma}
\affiliation{High Field Magnet Laboratory (HFML-EMFL), Radboud University, Toernooiveld 7, Nijmegen, 6525 ED, The Netherlands}
\affiliation{Radboud University, Institute for Molecules and Materials, Heyendaalseweg 135, 6525 AJ Nijmegen, The Netherlands.}	
\author{Lev Levitin}
\affiliation{Department of Physics, Royal Holloway University of London, Egham, TW20 0EX, Surrey, United Kingdom}
\author{Marijn Lucas}
\affiliation{Department of Physics, Royal Holloway University of London, Egham, TW20 0EX, Surrey, United Kingdom}
\author{Andrew Casey}
\affiliation{Department of Physics, Royal Holloway University of London, Egham, TW20 0EX, Surrey, United Kingdom}
\author{Jan Ny\'{e}ki} 
\affiliation{Department of Physics, Royal Holloway University of London, Egham, TW20 0EX, Surrey, United Kingdom}
\author{Ineke Broeders}
\affiliation{High Field Magnet Laboratory (HFML-EMFL), Radboud University, Toernooiveld 7, Nijmegen, 6525 ED, The Netherlands}
\author{Aaron Sutton}
\affiliation{Department of Physics, University of Toronto, 60 St. George Street, Toronto, M5S 1A7, Ontario, Canada}
\author{Bohdan Andraka}
\affiliation{Department of Physics, University of Florida, P.O. Box 118440, Gainesville, Florida 32611-8440, USA}
\author{Stephen Julian}
\affiliation{Department of Physics, University of Toronto, 60 St. George Street, Toronto, M5S 1A7, Ontario, Canada}
\author{John Saunders}
\affiliation{Department of Physics, Royal Holloway University of London, Egham, TW20 0EX, Surrey, United Kingdom}
\author{Alix McCollam}\affiliation{High Field Magnet Laboratory (HFML-EMFL), Radboud University, Toernooiveld 7, Nijmegen, 6525 ED, The Netherlands}
\email{alix.mccollam@ru.nl}


\email{}

\maketitle

%
The motivation to develop materials for quantum technologies 
has put exploration of novel quantum states of matter at the focus of 
several research fields, with particular efforts towards 
understanding and controlling the behaviour of quantum entangled 
and other strongly interacting electronic states~\cite{Osterloh2002, Hanson2007, BarGill2012, Kozii2016, Maryenko2018, Zaanen2019, Broholm2020, Soluyanov2015, Prochaska2020, Paschen2020}.
Experimental 
investigation is of primary importance, but requires measurements at 
ultra-low temperatures where the quantum states of interest have long 
lifetimes~\cite{Petta2005, Bluhm2010, Hanson2007, Pickett2018}. Under these conditions, low energy interactions, such as hyperfine or 
nuclear exchange interactions,
become relevant, and can modify electronic 
ground states and their associated excitations in multiple ways
that are not well understood or characterised \cite{Eisenlohr2021, Schuberth2016, Knapp2023, Bitko1996, Libersky2021, GimnezSantamarina2020}.
In this work, we 
use a recently developed magnetic susceptibility technique, compatible
with ultra-low temperatures and high magnetic fields \cite{ult_set-up_ref,Bangma_thesis}, to 
probe the influence of nuclear 
interactions on superconducting and multipole ordered ground states in 
the strongly correlated electron system \ce{PrOs4Sb12}.
We find that the multipole order develops a novel, 
entangled
nuclear-electronic character at the lowest temperatures, 
which significantly modifies the phase boundary and leads to 
a nuclear quantum critical point \cite{Ronnow2005, Eisenlohr2021}. 
In the superconducting phase, we find that hyperfine interactions 
suppress superconductivity in a manner that provides evidence for 
superconducting pairing mediated by crystal field excitations \cite{Fulde1970,Goremychkin2004,Matsumoto2004}. 
Our results on \ce{PrOs4Sb12}  
experimentally establish a new type of 
non-magnetic, nuclear quantum critical point,
and give 
revealing
insight into a highly
unusual superconducting state.
They also demonstrate more generally
the feasibility of exploiting
hyperfine interactions 
as a tuning parameter for experimental 
creation and investigation of a variety of quantum states and phenomena 
in correlated electron materials. 
%
%
\newpage

Strong correlations between electrons often lead to ordered phases in 
quantum materials that we can describe in terms of the electronic variables.
Nuclei provide extra degrees of freedom 
but are typically regarded independently of the electronic system because 
nuclear and electronic energy scales are well-separated.
There can be cases, however, where nuclear and electronic charge or spin 
couple  strongly, for example, through a large hyperfine interaction, 
and we must then consider how nuclear interactions affect electronic 
ordered phases and ground states.

To date, only a few examples have been studied where nuclear degrees of 
freedom notably enhance, 
suppress, or induce electronic order.
These examples, 
including \ce{YbRh2Si2}~\cite{Knapp2023, Nguyen2021, Schuberth2016}, \ce{LiHoF4}~\cite{Bitko1996}, 
\ce{PrCu2}~\cite{Andres1973} and \ce{Pr3Pd20Ge6}~\cite{Steinke2013, Iwakami2014}, 
cross several material families and involve diverse orders, 
such as superconductivity or magnetism. This shows the relevance of the 
effect in many contexts, but makes it difficult to extract general trends 
in behaviour.
Added to this is the fact that the low energy scales involved usually 
necessitate experiments at ultra-low
temperatures, which can be 
technically challenging as well as limited in dynamic range. 
Although temperatures below tens of 
millikelvin have been achievable for many decades, experimental probes, 
particularly thermodynamic probes, suitable for measurements of quantum 
solids in the ultra-low temperature regime have not been widely available.
This has blocked experimental access and limited progress in understanding
and characterising the influence of nuclear interactions on electronic 
ordered states.

\begin{figure*}[t]
	\centering
\includegraphics{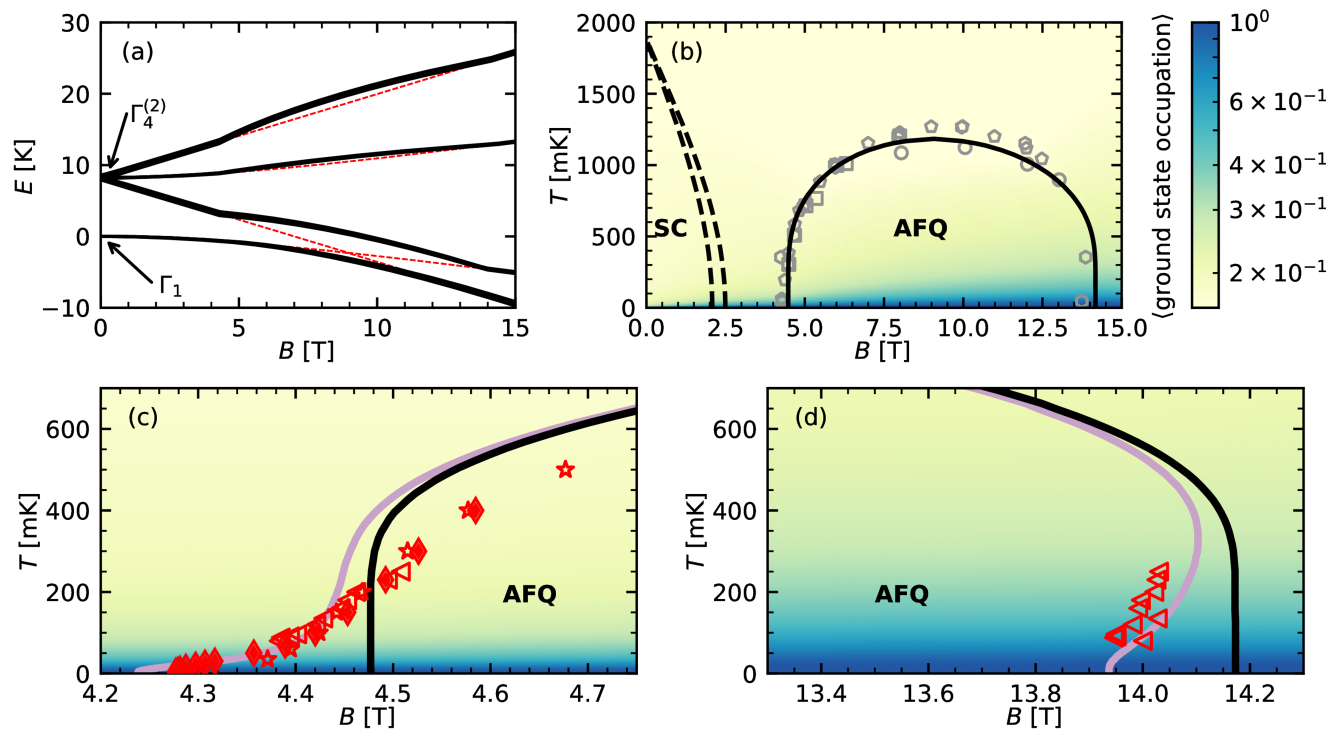}
	\caption{\textbf{Phase diagram and effect of hyperfine interactions for AFQ phase of \ce{PrOs4Sb12}.}
\textbf{(a)} Energy of the crystal-electric-field (CEF) levels as a function of magnetic field, in the presence (black lines) and absence (red lines) of quadrupole and hyperfine interactions. The energy splitting of the CEF levels in the presence of hyperfine interactions is too small to be resolved.
\textbf{(b)} Mean-field antiferroquadrupolar (AFQ) phase boundary excluding nuclear interactions (solid line). The dashed lines are the two upper critical fields of the superconducting phase in the absence of nuclear interactions, based on
Refs.~\cite{Bauer2002, Tayama2003, Measson2004}. Gray data points are from Aoki~(\textcolor{gray}{$\square$})~\cite{Aoki2002}, Tayama~(\textcolor{gray}{$\pentagon$})~\cite{Tayama2003}, Sugawara~(\textcolor{gray}{$\varhexagon$})~\cite{Sugawara2005}, and Rotundu~(\textcolor{gray}{$\ocircle$})~\cite{Rotundu2004}
\textbf{(c)} Measured lower and \textbf{(d)} upper AFQ phase boundary at low temperatures, from experiments at Royal Holloway~(\textcolor{red}{$\blacklozenge$}), Toronto~(\textcolor{red}{$\star$}) and Nijmegen~(\textcolor{red}{$\triangleleft$}). The black and light purple lines are the mean-field antiferroquadrupolar phase boundaries excluding and including nuclear interactions, respectively. 
 At low temperatures, the experimental data (and the light purple lines) show a dramatic departure from the mean-field phase boundaries calculated without hyperfine interactions. 
The colour maps in \textbf{(b)} to \textbf{(d)} show the
occupation probability of the (hyperfine) ground state. This level is fully and
solely occupied at the lowest temperatures, whereas it drops to $\sim \frac{1}{6}$ when the $T$ is a lot larger than the hyperfine constant. 
}
	\label{fig1}
\end{figure*}

To 
consistently address the question of nuclear-electronic coupling in quantum 
materials, we should ideally follow the evolution and influence of specific 
nuclear-electronic interactions across ordered phases and phase boundaries 
in a single material or closely related materials. 
This would require ultra-low temperature exploration of a system with multiple 
electronic orders that can be tuned via accessible parameters. 
We recently developed a high sensitivity magnetic susceptibility technique,
operating at temperatures down to $\sim 1$~mK in
magnetic fields up to $\sim 7$~T \cite{ult_set-up_ref}, which has allowed us 
to do this in the
strongly correlated electron material \ce{PrOs4Sb12}.
\ce{PrOs4Sb12} is an unconventional superconductor with $T_c$=1.85~K 
and $H_{c2}$=2.2~T~\cite{Bauer2002}. It develops magnetic field induced~\ac{afq} order 
between $\sim 4$ and 14~T~\cite{Aoki2002}, with low temperature phase boundaries that 
terminate in \acp{qcp}. (Fig.~\ref{fig1}(b)) shows the phase diagram of 
\ce{PrOs4Sb12} derived from experimental results (grey data points)  
and mean field calculations (solid black lines) [see Methods]. 
Previous work indicated that hyperfine interactions modify the \ac{afq} order 
below 300~mK
\cite{McCollam2013}, with the relatively high temperature scale being due 
to the combination of high nuclear spin (\textit{I}=5/2) and exceptionally 
large hyperfine constant ($A = 52$~mK) for Pr.
The colour map superimposed on Fig.~\ref{fig1} shows the occupation 
probability of the hyperfine ground state, such that the darkening colour 
indicates the increasing importance of hyperfine interactions. 
\ce{PrOs4Sb12} thus offers the opportunity to study how nuclear-electronic 
interactions affect superconductivity, multipolar order and quantum 
criticality.

We carried out magnetic susceptibility ($\chi$) measurements of 
\ce{PrOs4Sb12} in magnetic field up to 6~T at temperatures down to $\sim1$~mK,
and in magnetic field up to 30~T at temperatures down to $\sim 80$~mK 
(see Methods). 
Focussing first on the AFQ phase, we find that the phase boundaries 
continuously develop down to a few millikelvin, and are shifted to lower 
magnetic field below $\sim 300$~mK (Fig. 1 (c) and (d)).
We ascribe this behaviour to the influence of hyperfine interactions and 
the formation of a hybridised nuclear-electronic ground state. 
The Pr energy levels are 
no longer purely electronic 
entities in this region of the phase diagram, but must be described in terms 
of entangled nuclear and electronic states. 
The associated quantum critical points consequently have a fundamentally 
different character to their electronic counterparts,
and are examples of nuclear quantum critical points.

The superconducting phase is markedly suppressed at the lowest temperatures 
(Fig. \ref{fig3}(c)). In this case, we show that the behaviour is due to hyperfine enhanced magnetisation,
and gives new information about the superconducting pairing mechanism.

To understand the low temperature behaviour of \ce{PrOs4Sb12}, we must consider 
the Pr~4\textit{f}$^2$ electrons, which are subject to a tetragonal 
($T_h$) \ac{cef} from the surrounding Sb atoms. The crystal field splits the 
ninefold degenerate, spin-orbit coupled $J$=4 levels into a
non-magnetic singlet ground state $\Gamma_1$ ($E$~=~0~K) and an excited 
magnetic triplet $\Gamma_4^{(2)}$ ($E$~=~8~K), with additional \ac{cef} levels 
more than 130~K higher in energy. The system can therefore be treated as a 
two-level singlet-triplet case \cite{ShiinaAoki2004}.

In the superconducting state, the singlet-triplet crystal field excitation
has a dominantly quadrupole character, and this quadrupole exciton
is thought to be integral to the superconductivity 
\cite{Fulde1970, Goremychkin2004, Matsumoto2004, Zwicknagl2009}.

On application of a magnetic field, the ground state acquires a weak magnetic 
moment as the field mixes small amounts of $\Gamma_4^{(2)}$ into $\Gamma_1$~\cite{ShiinaAoki2004}. 
The $\Gamma_4^{(2)}$ triplet splits into three branches, such that $\Gamma_1$ 
and the lowest $\Gamma_4^{(2)}$ level move closer together with increasing 
field (see Fig 1(a)). Once these two lowest levels are sufficiently 
close in energy, the ground state becomes a 
superposition of $\Gamma_4^{(2)}$ and $\Gamma_1$, and \ac{afq} order develops.

The composition, occupation, and splitting of Pr \ac{cef} levels are thus 
fundamental to both the superconducting and \ac{afq} states.
It is precisely 
these properties that are strongly affected by the hyperfine interaction 
in \ce{PrOs4Sb12}, which 
enables the Pr nuclear spin to profoundly influence the low temperature 
phase diagram.
\\
%
%


\begin{figure*}[t]
	\centering
\includegraphics{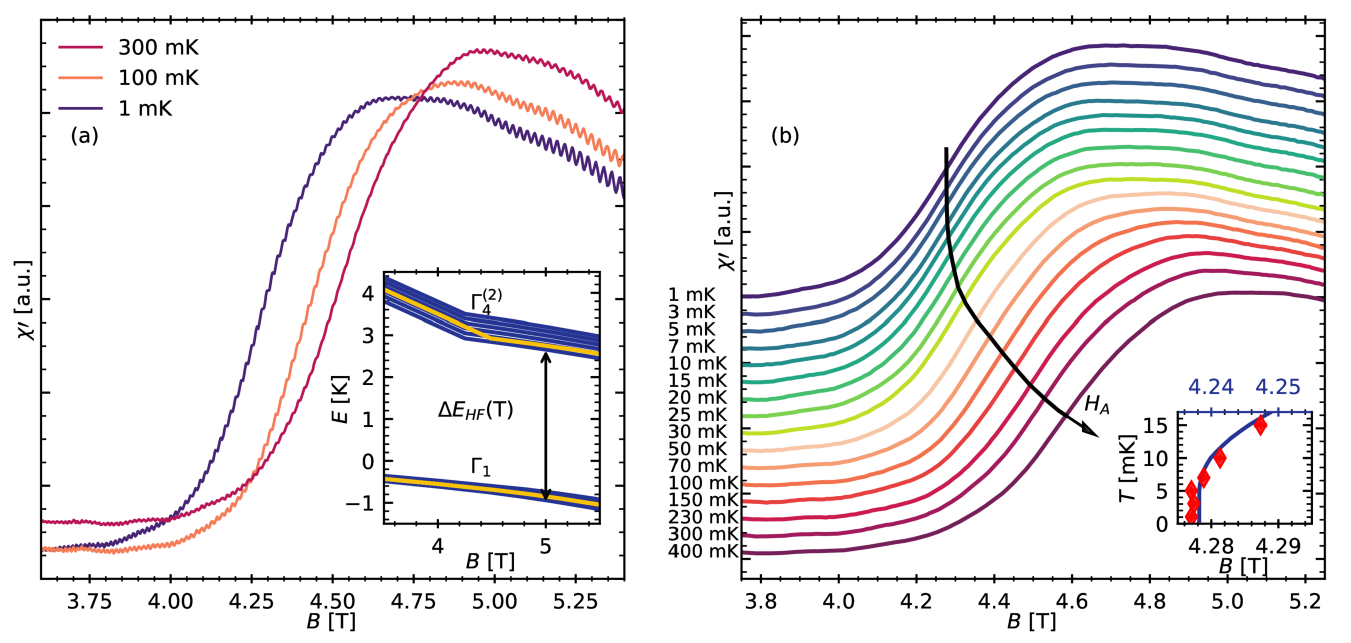}	
	\caption{\textbf{Hyperfine modified \ac{afq} phase boundary.}
		\textbf{(a)}~Real part of the magnetic susceptibility around the \ac{afq} transition at various temperatures. The step in $\chi^\prime(B)$ signals the \ac{afq} transition. Superimposed on this are quantum oscillations. The inset shows the mean field energy levels as a function of magnetic field excluding (yellow) and including (blue) nuclear interactions
		\textbf{(b)}~Real part of the magnetic susceptibility with quantum oscillations subtracted. Curves are vertically offset for clarity. The arrow indicates the evolution of the AFQ phase boundary down to the lowest temperatures. The inset shows the \ac{afq} phase boundary at ultra-low temperatures, saturating below 5~mK. Error bars in field are plotted, but are smaller than the data points. The blue line (top $x$-axis) is the lower AFQ phase boundary
calculated by mean-field theory including nuclear interactions.
	}
	\label{fig2}
\end{figure*}
%
%

%
%
\noindent \textbf{Modified \ac{afq} phase boundary}\vspace{0.1cm}\\
We first focus on the \ac{afq} phase transition, and in  
Fig.~\ref{fig2}(a) 
show the real part of the magnetic susceptibility $\chi^\prime(B)$ between 
3.6~T and 5.4~T, at different temperatures. $\chi^\prime (B)$ increases 
rapidly near 4~T, signalling the lower \ac{afq} phase transition, 
and shows de Haas-van Alphen oscillations at higher fields. 
Fig.~\ref{fig2}(b) shows a series of $\chi^\prime(B)$ curves with the 
quantum oscillations subtracted.  We extracted the phase boundary $H_A$ by 
differentiating  $\chi^\prime(B)$ and taking the field value corresponding 
to the peak of the derivative, which could be done more accurately when the 
modulations from the dHvA effect were removed.  $H_A$ {\em vs.} temperature is 
plotted as red data points in  Fig.~\ref{fig1}(c).
A different definition of $H_A$, for example, at the foot or shoulder of the 
transition, shifts the phase boundary but does not change its shape 
(see Methods and Extended Data Fig\ref{fig_SE:HA}).

Remarkably, $H_A$ develops strongly down to 5~mK, where it saturates within 
the experimental uncertainty, as shown in the inset to Fig.~\ref{fig2}(b). 
Our ultra-low temperature measurements could not be extended to the upper 
phase boundary near 14~T, but Fig~\ref{fig1}(d) shows that we observe a 
similar continued shift of the upper phase boundary to at least the lowest 
measured temperature of 80 mK.
This behaviour is
at odds with mean field models, which predict saturation of $H_A$ 
below $\sim 400$~mK \cite{Shiina2004, Kusunose2009}.
A mean field AFQ phase boundary (based on \cite{Bangma_thesis, ShiinaAoki2004, Shiina2004, Kusunose2009, Aoki2002, Tayama2003, Sugawara2005, Rotundu2004}) is shown as a 
solid black line in Fig~\ref{fig1}(b,c,d).

%
%
%
%
As discussed above, we expect nuclear degrees of freedom to be important 
below a few hundred millikelvin in \ce{PrOs4Sb12}. 
To explore the possibility that 
this is responsible for the anomalous development of the AFQ phase boundary,
we incorporated hyperfine interactions into the mean field model of the 
AFQ phase (see Methods).

Each \ac{cef} level is a manifold of (2$I$+1)=6~nuclear levels. 
The hyperfine interaction lifts their degeneracy, and causes an energy 
splitting proportional to their magnetic moment. This is illustrated in the 
inset of Fig.~\ref{fig2}(a)), which shows a significantly larger splitting 
in the magnetic $\Gamma_4^{(2)}$ state than in the barely magnetic ground 
state, which consists primarily of $\Gamma_1$.
As a result, the energy gap from $\Gamma_1$ to $\Gamma_4^{(2)}$ diminishes 
as the system is cooled and occupation of the lowest hyperfine level increases. 
Equivalently, we can think of this
as the hyperfine Hamiltonian mixing an increasing amount of $\Gamma_4^{(2)}$ 
into $\Gamma_1$ as temperature is reduced. 
The AFQ order is thus stabilised at increasingly lower magnetic field,
an effect which saturates 
when the lowest hyperfine level is fully occupied.  

The light purple line in Fig.~\ref{fig1}(c and d) represents the mean field AFQ phase boundary when the nuclear interactions are included. It is in good qualitative agreement with the experimental $H_A$ (red data points), as it captures the strong shift to lower field as the temperature is reduced. The saturation of the lower phase boundary below 5~mK is also well-described by the model (see inset of Fig.~\ref{fig2}(b)), indicating that the system enters its quantum ground state below 5 mK.

The overall shape of the mean field AFQ phase boundary is not a good quantitative match to the experimental data on the scale of a few hundred millikelvin. This can be clearly seen in Fig.s~\ref{fig1}(b) and (c): on the 2~K and 15~T scale of (b), the data points appear to match well to the mean-field AFQ phase boundary, however, systematic deviations on the scale of 100~mK/0.1~T can be clearly seen in Fig.~\ref{fig1}(c).
(See Methods for further discussion.)
Even by adjusting the parameters in the mean field model, we are not able to accurately match the calculated phase boundary to the data at all fields and temperatures. 
This is an important finding, because it shows that although a mean field picture is 
extremely useful for broadly mapping and understanding the phase diagram of \ce{PrOs4Sb12}, it does not give a quantitative description at the energy scale relevant for the hyperfine interactions and lowest temperature 
behaviour.\\

%
%
%
%
%
%
%
%
%
\noindent
\textbf{Nuclear quantum criticality}\\
We have shown that the AFQ order in \ce{PrOs4Sb12} is strongly 
temperature dependent in the composition of the states that order, as well as in its phase boundary. 
At high temperatures, the quadrupoles are formed on the (electronic) Pr crystal field levels, %
 but at temperatures below $\sim$ 300~mK they develop on entangled crystal field and nuclear states.
The low temperature AFQ phase is thus an example of hybrid nuclear-electronic order, and the corresponding quantum critical points are 
of a novel type,
associated with transitions between disordered and ordered nuclear-electronic phases.

At a continuous quantum phase transition, the energy of collective electronic
excitations is expected to go to zero at the quantum critical point~\cite{SachdevBook}.
In systems with significant hyperfine coupling, electronic and nuclear modes 
mix, such that there are two sets of dispersive modes, 
one dominantly electronic, and one at lower energy that is dominantly 
nuclear. 
At the 
quantum critical point, 
softening of the dominantly electronic mode is
incomplete and it is the nuclear mode that becomes gapless. This behaviour 
leads to  
nuclear quantum criticality, which 
cuts off electronic criticality below a certain temperature 
that depends
on the hyperfine coupling strength \cite{Eisenlohr2021, Libersky2021, McKenzie2018, Kovacevic2016,Ronnow2005}

The AFQ quantum critical points in \ce{PrOs4Sb12} are nuclear quantum critical points, and provide
a rare opportunity to explore and characterise nuclear quantum criticality.
The only previous extended investigation of nuclear quantum criticality  
has been in 
the transverse-field Ising magnet LiHoF$_4$ \cite{Ronnow2005,Ronnow2007, Kovacevic2016, McKenzie2018}, although the effect of nuclear spins on a number of other model magnetic quantum phase transitions was discussed in recent theoretical work~\cite{Eisenlohr2021}.
A non-magnetic nuclear quantum critical point, such as the AFQ nuclear quantum critical point we observe in PrOs4Sb12, has not previously been studied either
experimentally or theoretically. 

In \ce{LiHoF4}, the transfer of quantum criticality from the electronic to the nuclear mode
is accompanied by a marked shift of the quantum critical point~\cite{Bitko1996}, due to stabilisation of the magnetic phase by interaction with the nuclear spins. 
The modified AFQ phase boundary we observe in \ce{PrOs4Sb12} suggests that 
the general behaviour on approaching the nuclear QCP is similar to that 
in \ce{LiHoF4}. 
However, the 
different role of nuclear spins in a quadrupolar system 
is likley to be important for a quantitative understanding.
Our results on \ce{PrOs4Sb12} strongly motivate theoretical work and further experiments to 
examine nuclear quantum criticality 
associated with multipole and other non-magnetic order parameters.
It is of particular interest to examine the crossover with decreasing temperature between the regimes of electronic and nuclear quantum criticality. The behaviour near nuclear quantum criticality in \ce{LiHoF4} has been discussed in terms of interaction of the Ising system with a spin bath, and an interesting region of minimum decoherence of the 2-level spin system was proposed in this crossover region, just at the temperature where the ferromagnetic phase boundary begins to be modfied by the hyperfine interactions~\cite{Ronnow2005}.\\
\begin{figure*}[t]
	\centering
\includegraphics[trim=2 0 1 0,clip]{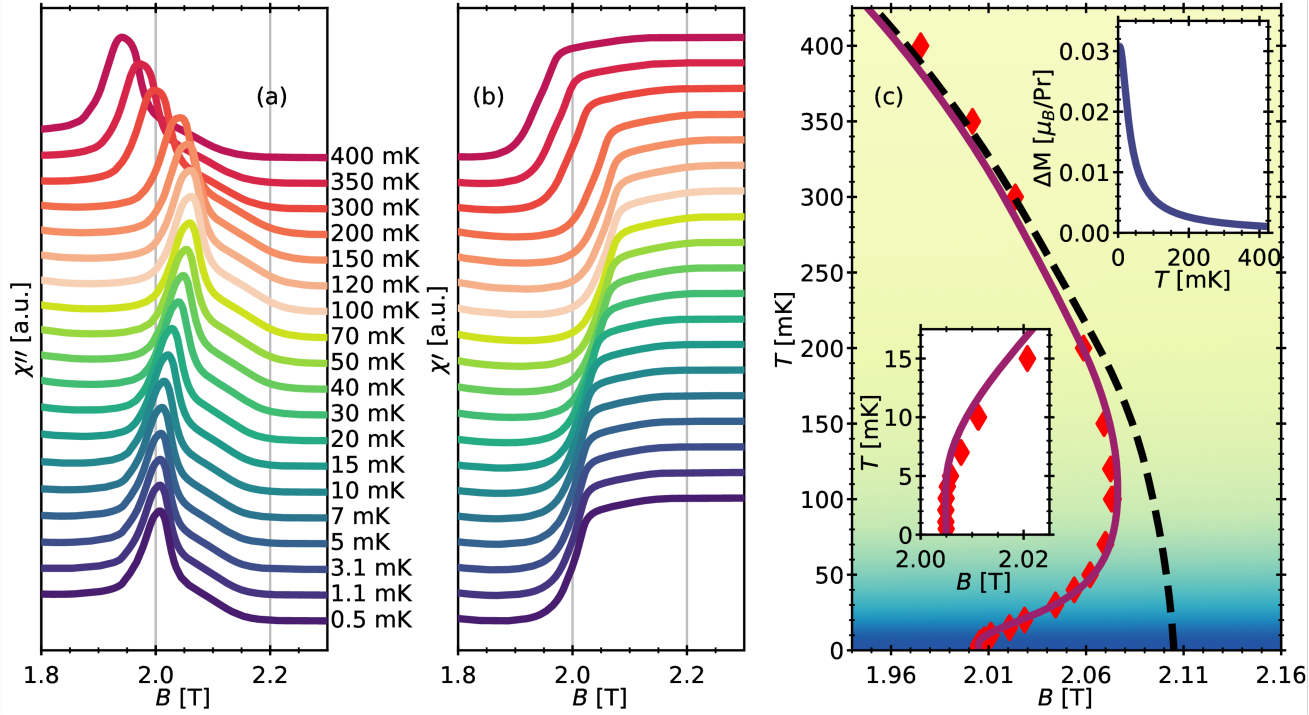}
	\caption{\textbf{Reduction of the superconducting upper critical field by hyperfine interactions.} \textbf{(a)}~Imaginary and \textbf{(b)}~real parts of the magnetic susceptibility at various temperatures. The peak and step respectively signal the superconducting transition. The curves are vertically offset for clarity. \textbf{(c)}~Comparison between the extracted $H_{c2}$(T) from (a,b) (\textcolor{red}{$\blacklozenge$}) and calculated $H_{c2}$ (purple line) based on the model described in the text. The dashed, black line shows $H_{c2}$ of the model in the absence of nuclear interactions. The right inset shows the additional magnetization at very low temperature caused by nuclear interactions, as calculated by mean field theory. The left inset shows $H_{c2}$ at the lowest temperatures.}
	\label{fig3}
\end{figure*}
%
%
\noindent
\textbf{Hyperfine-induced suppression of SC}\\
We turn now to the superconducting region of the phase diagram. Figure~\ref{fig3} shows the imaginary~(a) and real~(b) magnetic susceptibility crossing the superconducting transition down to 5~mK. 
The sharp peak in $\chi^{\prime\prime}$ and step in $\chi^\prime$ signal the upper critical field, $H_{c2}$. The slight shoulder visible in $\chi^{\prime\prime}$ at higher field is the second superconducting transition that is typically observed in \ce{PrOs4Sb12} \cite{Bauer2002, Measson2008, Andraka2012}. 
We focus on the temperature dependence of $H_{c2}$ below 400~mK, shown in Fig.~\ref{fig3}(c). Interestingly, we observe a strong decrease of $H_{c2}$ below 100~mK, until it eventually saturates below 2~mK.

The mechanism of superconductivity in \ce{PrOs4Sb12} is not fully established. 
The sister compound \ce{LaOs4Sb12} is understood to be a multiband superconductor, with conventional, phonon-mediated pairing on two bands, and $T_c = 740$~mK \cite{Tee2012}. The significantly higher $T_c$ in \ce{PrOs4Sb12} is therefore clearly linked to the presence of Pr $4f$ electrons, and is thought to involve an unconventional pairing channel on one of the bands \cite{Seyfarth2005}. 
A number of scenarios have been proposed. One scenario is that a strong interaction of the conduction electrons with the Pr $4f$ charge density 
through inelastic charge scattering promotes Cooper pair formation. This process involves exciting quadrupolar excitons (dispersive excitations of the \ac{cef} levels)~\cite{Fulde1970, Sugawara2002, Goremychkin2004, Koga2006, Chang2007, Namiki2007, Sato2010}, which could play a similar role to phonons in conventional BCS theory.
The inelastic charge scattering competes with, but can dominate the pair-breaking 4\textit{f}--conduction electron exchange interaction.
The absence of a magnetic moment in $\Gamma_1$ and the quadrupolar character of the exciton in \ce{PrOs4Sb12} provides optimal conditions to realize this.

Another proposal is superconducting pairing mediated by quadrupole fluctuations associated with the nearby \ac{afq} order and quantum critical point, similar to spin-fluctuation mediated superconductivity in the vicinity of a magnetic quantum critical point  
~\cite{Cox1995, Mathur1998,Chia2003, Aoki2003, Aoki2007}.
However, increased quadrupolar susceptibility 
is reported
to be associated with suppression of superconductivity in \ce{PrOs4Sb12}, rather than enhancement~\cite{Tayama2006}, which is contrary to the scenario of quadrupole fluctuation mediated pairing. 
%
%
%
We observed that the \ac{afq} quantum critical point moves to lower magnetic field as a result of interaction with the nuclear states, which increases the quadrupole fluctuations at low temperature in the region of superconductivity~\cite{Eisenlohr2021}. 
Calculation of the 
mean field quadrupolar susceptibility at 2~T shows that it rises steeply below $\sim 100$~mK [see Methods and Extended Data Fig. \ref{fig_SE:alternative_Hc2_mech}), but we observe suppression of the superconductivity in this temperature range. This suggests that quadrupole fluctutations directly compete with superconductivity.

The similar temperature scale of the modified superconducting and AFQ phase boundaries in \ce{PrOs4Sb12} points to a 
similar origin.
We therefore look further into the role of nuclear interactions in the 
superconducting state. 

Independently of the superconducting or AFQ orders, the hyperfine interaction leads to a strongly increasing magnetic moment on the Pr atoms below $\sim$200~mK, due to the mixing of $\Gamma_4^{(2)}$ and $\Gamma_1$ described above.
The change in (mean field) magnetisation $\Delta M$ per Pr atom, as a result of this mixing, is shown as a function of temperature in the inset of Fig.~\ref{fig3}(c)). 
The conduction electrons interact with the 4\textit{f} magnetic moment through the exchange interaction, which tends to break Cooper pairs. In this way, the hyperfine interaction leads to suppression of superconductivity through enhanced exchange scattering.

We model the effect of this exchange scattering by using the generalized Abrikosov-Gorkov theory of Fulde and Maki \cite{FuldeMaki1966}.
This theory describes how magnetic impurities reduce $H_{c2}$ by creating an internal field and enhancing exchange scattering.
The internal field set-up by the Pr atoms is negligible, and the role of the magnetic impurities in our case is played by the hyperfine-enhanced magnetic moment, i.e. $\Delta M$. 
We use the mean-field $\Delta M$ to capture the amount of exchange scattering:
\begin{equation}\label{eqn:Fulde}
	\begin{split}
		& H_{c2}(T, A=\textrm{52~mK}) = \\
		& H_{c2}(T, A=\textrm{0}) - H_{c2}(T=0, A=0)\frac{\Delta M (T)}{M_{\textrm{crit}}}
	\end{split}
\end{equation}
$H_{c2}(T, A=0)$ is the upper critical field in the absence of nuclear interactions, for which we used a curve based on high-temperature data points~\cite{Maki2007}, above the temperatures relevant for hyperfine interactions.  $M_{\textrm{crit}} = 0.83~\mu_B$/Pr was set to ensure a match with our lowest temperature data point (see Methods for details). 

$H_{c2}(T)$ calculated from this model is shown as a purple line in Fig.~\ref{fig3}(c). The lower inset shows an expanded view of the lowest temperature region. The calculated curve adequately describes the experimental data (red points):
The maximum $H_{c2}(T)$ of the model is in very close agreement with the experimentally observed maximum at 100~mK, and the saturating $M$ of the Pr atom below 2~mK explains the observed saturation of $H_{c2}$ below 2~mK. Thus, we conclude that superconductivity in \ce{PrOs4Sb12} is suppressed by the hyperfine-enhanced magnetisation of Pr as it enhances pair-breaking exchange scattering.

In addition to enhanced quadrupole fluctuations and enhanced exchange scattering, we have also considered the effect of the Pauli limiting field and orbital limiting field on $H_{c2}$.
We find that Pauli limiting would lead to too small a change in $H_{c2}$ to explain the observed behaviour, and orbital limiting would increase, rather than decrease, $H_{c2}$ (see Methods and Extended Data Fig. \ref{fig_SE:alternative_Hc2_mech}). 

Limited exchange scattering between 4$f$ and conduction electrons is a prerequisite for superconductivity mediated by quadrupolar excitons \cite{Fulde1970}, where the balance between exchange scattering and inelastic charge scattering determines the superconducting $T_c$.
The observed sensitivity of superconductivity to even weakly enhanced 
exchange scattering indicates the importance of this balance in 
\ce{PrOs4Sb12}. Our ultra-low temperature results therefore strongly suggest that 
Cooper pairing 
via inelastic charge scattering and the formation of quadrupolar excitons  
contributes significantly to superconductivity in \ce{PrOs4Sb12}.
This quadrupolar exciton mechanism should be in addition to  phonon-mediated pairing, comparable to that observed in \ce{LaOs4Sb12}, and can account for the increased $T_c$ in the Pr compound.

Note that in the above we have focused on the first (lower field) superconducting transition because it is the dominant feature in the susceptibility. 
However, as can be seen in  Fig.~\ref{fig3}(a), the higher field shoulder in $\chi^{\prime\prime}$, representing the second superconducting transition, shows the same temperature dependence as the first transition, and the two $H_{c2}$ phase boundaries have the same shape. This implies that both superconducting transitions are affected in the same way by hyperfine interactions. 
The second superconducting transition in \ce{PrOs4Sb12} is robust in samples above a certain size~\cite{Andraka2012}, and the double transition has been proposed as evidence of non-unitary chiral superconductivity in this material~\cite{Kozii2016}.
If the two transitions delineate distinct superconducting phases~\cite{Kozii2016}, our results suggest that the superconducting mechanisms in the two phases are similar. \\

%

%
\noindent
\textbf{Summary and concluding remarks}
\\
Development of a magnetic susceptibility technique 
compatible with ultra-low temperature~\cite{ult_set-up_ref, Bangma_thesis} has allowed us to 
sensitively probe the ground state of \ce{PrOs4Sb12} and examine the 
effect of nuclear degrees of freedom 
on electronic ordered phases. We find that hyperfine interactions have 
a strong effect on both the AFQ and superconducting states, but that 
the consequences for the electronic order are quite different in each case.

In the AFQ phase, the mixing of electronic and nuclear modes by the hyperfine interaction leads to a novel, hybrid nuclear-electronic AFQ order below $\sim 300$~mK, with a strongly modified phase boundary and a new type of nuclear AFQ quantum critical point. 
Our experiments have provided the first access to nuclear quantum criticality
in the ultra-low temperature regime, where we have 
been able to measure the saturation of the nuclear-electronic system into the 
hyperfine ground state below 5~mK.
\ce{PrOs4Sb12} is thus an ideal system in which to further explore 
the development of nuclear quantum criticality from electronic quantum 
criticality, and may be of particular interest in the contexts of coherent nuclear-electronic entangled states~\cite{Ronnow2005, Stamp2004, Fraval2004}, and non-Fermi liquid behaviour associated with AFQ quantum criticality~\cite{Patri2020,Han2022}.

In the superconducting phase, we find that $H_{c2}$ is 
suppressed at temperatures below $\sim 200$~mK, in a way that 
is well-described by a model of hyperfine-enhanced exchange scattering.
The extreme sensitivity of the the superconducting state to exchange 
scattering in this material suggests a quadrupole exciton mediated mechanism of superconducting pairing~\cite{Fulde1970, Goremychkin2004}.
This new information about the likely pairing mechanism in \ce{PrOs4Sb12} goes some way to resolving long-standing questions about the nature of superconductivity in this material. Further work to understand if two-band superconductivity partly mediated by quadrupole excitons is compatible with proposed non-unitary chiral superconductivity~\cite{Kozii2016} would be valuable.

The results we present in this article demonstrate both the feasibility and the value of thermodynamic measurements on quantum solids at ultra-low temperatures and high magnetic fields. We observe the strong influence of hyperfine interactions on low temperature electronic orders, and gain striking new information about the nature of quantum criticality, the AFQ ground state and superconductivity in \ce{PrOs4Sb12}. 
More generally, this work provides clear evidence that, as we search for materials to support quantum technologies, experiments of this kind can contribute key information about the role of nuclear degrees of freedom in phenomena such as quantum order and quantum criticality, which shape the low temperature behaviour and stability of potentially important quantum states~\cite{Pickett2018}.\\\\

\noindent
\textbf{Acknowledgements}\\
We thank Lijnis Nelemans, Michel Peters, Richard Elsom, Paul Bamford and Ian Higgs for excellent mechanical support.
The research leading to these results has received funding from the European Union’s Horizon 2020 Research and Innovation Programme, under Grant Agreement no 824109 (European Microkelvin Platform). Part of this work was also supported
by HFML-RU/NWO-I, a member of the European Magnetic Field Laboratory (EMFL).\\




\newpage

\noindent
\section*{Methods}
\setcounter{page}{1}
\setcounter{figure}{0}
\setcounter{table}{0}
\setcounter{equation}{0}
\renewcommand{\figurename}{Extended Data Fig.}

\noindent
{\bf Sample growth}\\
Single crystals of \ce{PrOs4Sb12} were grown by a standard Sb-self-flux method~\cite{Bauer2001}. The sample measured at the High Field Magnet Laboratory (HFML) and at Royal Holloway University of London (RHUL) had dimensions 0.9~$\times$~0.95~$\times$~1.05~mm$^3$, and the sample measured at the University of Toronto had dimensions 2.5~$\times$~2~$\times$~2~mm$^3$. The samples had similar residual resistance ratios between 70 and 80. 
\\\\
{\bf Experiments}\\
Magnetic susceptibility was measured using the field-modulation technique, with the sample placed in one of two counterwound pick-up coils, mounted in the center of a modulation coil and a larger sample magnet.
The lowest-temperature experiments were carried out at the Department of Physics, RHUL. Measurements were performed in a nuclear demagnetization refrigerator with a Cu nuclear stage and 9~T superconducting sample magnet.
The data shown in the main paper were acquired in a modulation field of 1.4~G at a frequency of 92.34~Hz. To investigate possible sample heating due to the modulation field, measurements were repeated at reduced modulation amplitudes of 0.03~G, 0.14~G and 0.55~G at the lowest temperatures (see {\bf Sample temperature} below). The voltage across the pick-up coils was amplified by a temperature-stabilized and lead-shielded low-temperature transformer, and measured with a lock-in amplifier. More detailed information about the experimental set-up is given elsewhere~\cite{ult_set-up_ref}.
Data were obtained with the sample stabilized at a constant temperature, while ramping the sample magnet from high to low field at $-0.05$~mT/s. 
To prevent hysteretic effects, the system was always stabilized at the required temperature before entering the \ac{afq} phase by magnetisation across $H_A$.

\begin{figure*}[h]
	\centering
	\includegraphics{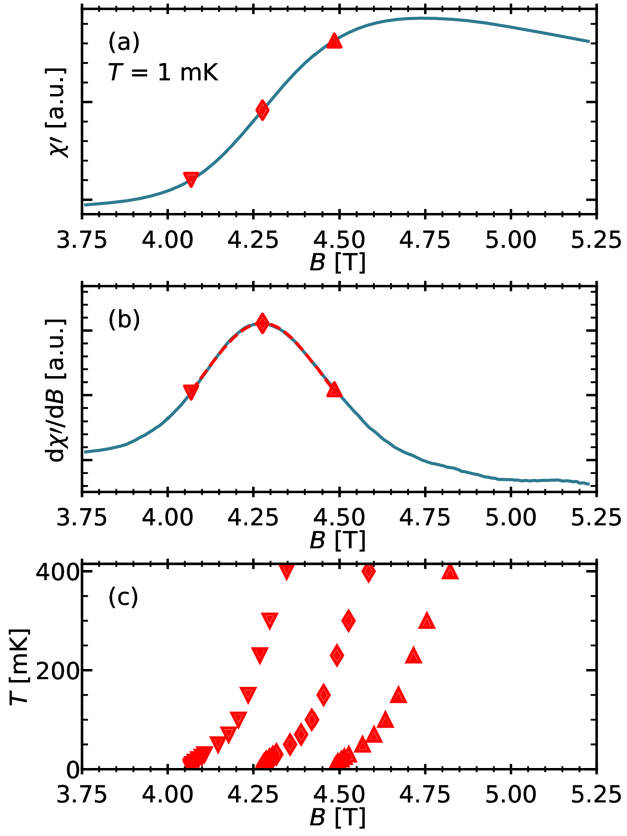}	
	\caption{ 
\textbf{Extraction of lower \ac{afq} phase boundary.} (a)~$\chi^\prime$ at 1~mK from which quantum oscillations have been subtracted. (b)~Field-derivative of $\chi^\prime$ from (a) with a fit to a Gaussian function (red dashed line). The diamond indicates the center of the Gaussian fit, and the downward and upward triangles represent the half-height points of the same Gaussian at low and high field, respectively, and are also shown in (a). The triangles are close to the foot and shoulder of the transition. (c)~Temperature dependence of the \ac{afq} phase boundary based on the center (diamonds) or low/high field half-heights (downwards/upwards triangles) of a Gaussian fit. The different definitions shift the phase boundary, but do not qualitatively change its shape.}
\label{fig_SE:HA}
\end{figure*}

Magnetic susceptibility measurements at the HFML and University of Toronto were carried out using regular dilution refrigerators to base temperatures of 88~mK and 30~mK, respectively. The field-modulation technique was used in both cases.
\\\\
{\bf Determination of the AFQ phase boundary}\\
We extracted the \ac{afq} phase boundary $H_A$ by differentiating $\chi^\prime(B)$ and taking the field value corresponding to the peak of the derivative, which could be done more accurately when the modulations from the de Haas--van Alphen effect were removed.
A different definition of $H_A$ shifts the phase boundary but does not qualitatively change its shape.
For example, if we take the foot or the shoulder of the transition, defined to be at the low- or high-field half-heights of Gaussian fits to $\text{d}\chi^\prime/\text{d}B$, we obtain the phase boundary shown in Extended Data Fig.~\ref{fig_SE:HA}.
\\\\
{\bf Sample temperature}\\
Noise thermometry~\cite{Casey2014} and melting-curve thermometry~\cite{Greywall1985} were used during the lowest temperature measurements (RHUL). Both thermometers were located outside of the high field region, at a distance of $\sim$ 70 cm from the experiment. To obtain an estimate of the thermal gradients in the system, an extensive thermal model was developed and tested, details are given in ref.~\cite{ult_set-up_ref}.
The main source of heat close to the sample is eddy current heating due to the modulation field. Repeated measurements of the \ac{afq} phase boundary at ultra-low temperatures show no difference when performed at 0.6~G or 1.4~G, and thus give no indication of heating by eddy currents in the vicinity of the \ac{afq} phase. In addition, the saturation of the \ac{afq} phase boundary below 5~mK is in good agreement with mean-field models, which predict saturation below 7~mK. 
We therefore conclude that there is no significant eddy current heating of the sample at the modulation fields used.
\\\\
{\bf Sample temperature in the superconducting phase and determination of ${\bf H_{c­2}}$}\\
As discussed above, no heating of the sample was observed in the \ac{afq} phase. When the sample is in the superconducting phase, however, it is subject to additional heating due to the motion of flux vortices. The vortex motion is driven by both the modulation field and by sweeping the DC field, and is a function of the penetration field ($B_p$)~\cite{Bean1964}. $B_p$ scales linearly with the critical current density $j_c(B)$. We therefore expect this heating to be insignificant close to $H_{c2}$, where $j_c$ is small, and to grow as we enter further into the superconducting phase. Here we show that this heating is negligible very close to the normal state.

\begin{figure*}[t]
	\centering
	\includegraphics{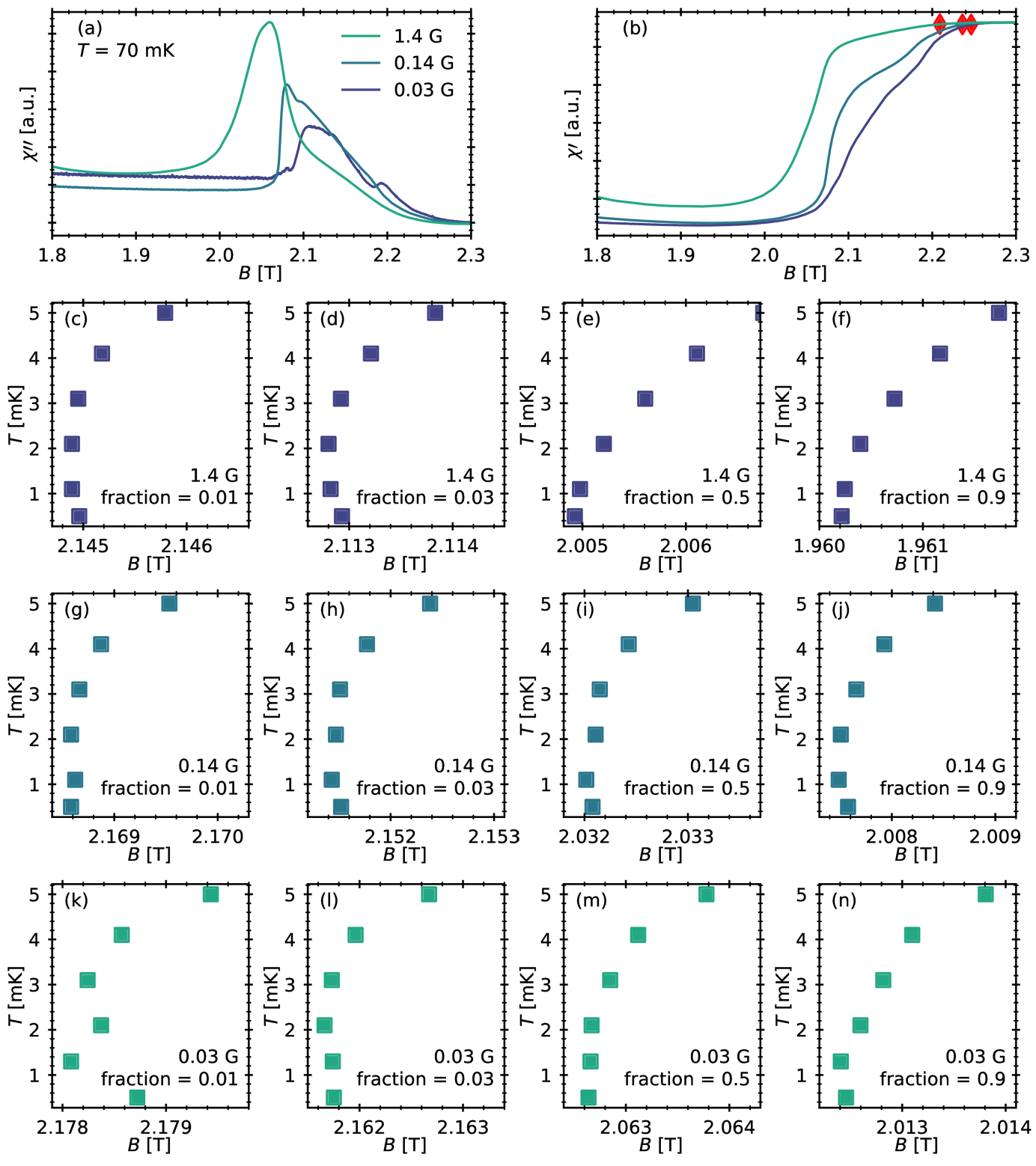}
	\caption{
\textbf{Superconducting transition at various modulation fields for different definitions of $\bf{H_{c2}}$.} Real~\textbf{(a)} and imaginary~\textbf{(b)} magnetic susceptibility at 70~mK. Data are normalized to the amplitude of the modulation field. A smaller modulation field moves the step in $\chi\prime$ and peak in $\chi\prime\prime$ to higher magnetic field~(e.g. \cite{Ge2014, Mller1989}). The resulting different upper critical fields are indicated by red diamonds and are defined at 1\% of the step down in $\chi^{\prime}$ (grey, dashed lines for 1.4~G data).  (c)~-~(n): $H_{c2}(T)$ for different modulation fields of 1.4~G ((c)~-~(f)); 0.14~G ((g)~-~(j)); and 0.03~G ((k)~-~(n)), defined at fractions of 1\%, 3\%, 50\% and 90\% of the step down in $\chi^{\prime}$. See text.
}
\label{fig_SE:SC_drives}
\end{figure*}

The superconducting phase transition is recognized in the AC susceptibility as a step in $\chi^{\prime}$ and a peak in $\chi^{\prime\prime}$, see Extended Data Fig.~\ref{fig_SE:SC_drives}(a),~(b). We define $H_{c2}$ as some fraction of the step down in $\chi^\prime(B)$. Choosing different fractions leads to small apparent shifts in $H_{c2}$ (see Extended Data Fig.\ref{fig_SE:SC_drives}). Smaller apparent values of $H_{c2}$ (larger fractions) are further into the superconducting state and imply larger values of $j_c$ and associated vortex heat.

Based on mean-field calculations, we expect to fully occupy the quantum ground state of \ce{PrOs4Sb12} at 2~T in the low-mK temperature range (see discussion and \textcolor{blue}{Fig. 3(c)} in the main paper). In this regime, saturation of $H_{c2}$ should occur. At higher temperatures, changing occupations of the hyperfine levels will result in temperature dependence of $H_{c2}$ (see discussion in main paper).
In Extended Data Fig.~\ref{fig_SE:SC_drives}(c)-(f) we compare the temperature dependence of $H_{c2}$ between 0.5 mK and 5 mK obtained at $b_{mod}$~=~1.4~G.
For $H_{c2}$ defined at fractions of 0.01 and 0.03 of the step down in $\chi^\prime(B)$ (Extended Data Fig.~\ref{fig_SE:SC_drives}(c)~-~(d)), $H_{c2}$ indeed shows saturation below 3~mK. This behaviour is also observed in measurements obtained
at lower modulation fields of 0.03~G and 0.14~G (Extended data Fig.~\ref{fig_SE:SC_drives}(g)-(n)), and indicates that vortex motion does not heat the sample. Definitions of $H_{c2}$ with significantly higher fractions, e.g. 0.5 and 0.9 in Extended Data Fig.~\ref{fig_SE:SC_drives}(e)~-~(f), which probe the sample further into the superconducting state, show a continuously evolving $H_{c2}$ down to the lowest temperatures, indicating heating
We therefore defined the upper critical field $H_{c2}$, as shown in Fig.~\ref{fig3}, as 1\% of the step down in $\chi^\prime(B)$, where we know the sample is not subject to large heating, see Extended Data Fig.~\ref{fig_SE:SC_drives}(b).
\\\\
{\bf Understanding the superconducting upper critical field boundary in PrOs$_4$Sb$_{12}$}\\
To understand the anomalous behaviour of the upper critical field boundary in \ce{PrOs4Sb12}, we use a model derived from the generalized Abrikosov-Gorkov theory of Fulde and Maki \cite{FuldeMaki1966}, which we modify to make applicable to \ce{PrOs4Sb12}.
The original model describes how magnetic impurities reduce $H_{c2}$ via (1)~spin-flip scattering of conduction electrons by magnetic impurities and (2)~spin polarization of conduction electrons by the exchange field generated by magnetic impurities:
\begin{equation}\label{eqn:original_Maki_Fulde}
\begin{split}
&H_{c2}(x, T) = \\
&H_{c2}(0, T) - H_{c2}(0, 0)\Big( \frac{x}{x_{crit}} + \frac{P}{P_{crit}}\Big)
\end{split}
\end{equation}
in which $x$ is the concentration of magnetic impurities, $H_{c2}(x, T)$ is the upper critical field at temperature $T$, and $P$ is the Pauli polarization term.~\cite{FuldeMaki1966}
\\
In the case of \ce{PrOs4Sb12}, we assume that the nuclear spins act as magnetic impurities. As limited exchange scattering is considered to be key for superconductivity in \ce{PrOs4Sb12}, we limit ourselves to pair-breaking mechanism (1) above.
In adapting the Fulde-Maki model, we convert the concentration of magnetic impurities $x$ to an increase in magnetization $\Delta M$, caused by the hyperfine interaction of strength $A$. Our expression for the upper critical field is then:

\begin{equation}
\begin{split}
& H_{c2}(A=52~\textrm{mK}, T) = \\
& H_{c2}(0, T) - H_{c2}(0, 0)\Big( \frac{\Delta M (T)}{M_{crit}}\Big)
\end{split}
\end{equation}
in which $\Delta M (T) = M(A=52~\textrm{mK}, T)-M(A=0, T)$ is the difference in mean-field calculated magnetizations including ($A=52$~mK) and excluding ($A=0$) the hyperfine interaction. $\Delta M$ is expressed in $\mu_B/\textrm{Pr atom}$ by comparison to measured magnetization~\cite{Tayama2003}.
$H_{c2}(A=0, T)$ is taken from Maki~\textit{et al}.~\cite{Maki2007}, based on measurements of \ce{PrOs4Sb12} at high temperature, which are largely independent of effects related to Pr nuclei.
The $H_{c2}$ values in Maki \textit{et al.} were larger than the values we measured. This could be due to differences in sample quality or because of a different definition of $H_{c2}$. We therefore rescaled $H_{c2}(A=0, T)$ from \cite{Maki2007} to match our high temperature data points above 300~mK. This results in a value of $H_{c2}(A=0, T=0)=2.245$~T.
$M_{\textrm{crit}} = 0.68~\mu_B$/Pr was set to ensure a match with our lowest temperature data point at 1~mK.

Enhanced exchange scattering is not the only mechanism that could suppress superconductivity. We also considered how hyperfine effects on Pauli limiting or orbital limiting could modify $H_{c2}$.

In the case of Pauli limiting, we convert $\Delta M$, determined above, to an internal field. When this is added to the external applied field, we find that the Pauli-limiting field changes by $\sim$~1~mT. This change is orders of magnitude too small to account for the observed suppression of $H_{c2}$.

\begin{figure*}[t]
	\centering
	\includegraphics{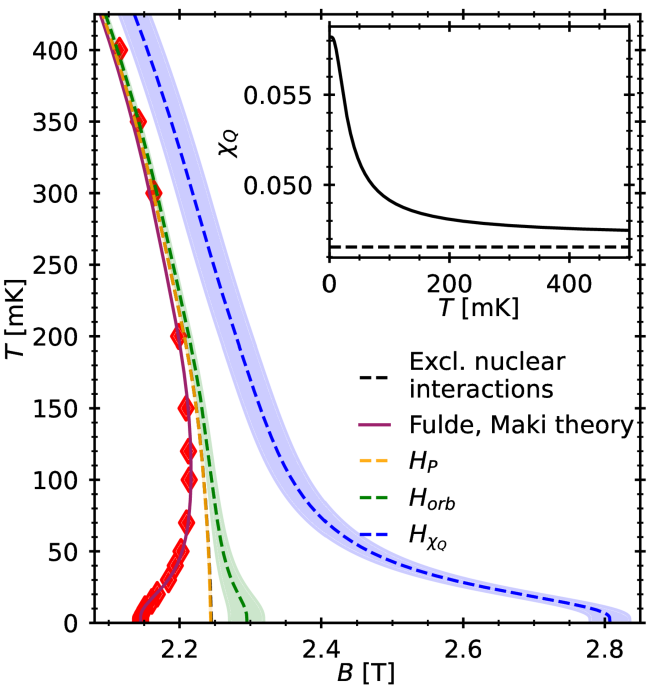}	
	\caption{\textbf{Expected changes in the upper critical field from various mechanisms due to hyperfine interactions.} $H_{\textrm{orb}}$ and $H_{{\chi}_Q}$ are estimates, with shaded areas added to give an idea of the uncertainty in the estimate. The inset shows the quadrupolar susceptibility vs temperature, expression~\ref{quad_susc} \cite{Zwicknagl2009}.} 
\label{fig_SE:alternative_Hc2_mech}
\end{figure*}

We calculate the effect on the orbital-limiting field by considering the temperature dependence of the effective mass $m_e$ of the quasiparticles: $H_{\text{orb}}\propto (1/\xi^2) \propto m_e^2$. A suggested mass-enhancement mechanism of quasiparticles in \ce{PrOs4Sb12} is through the aspherical coulomb scattering, captured in terms of the quadrupolar susceptibility~\cite{Zwicknagl2009}:

\begin{equation}
\frac{m_e}{m_0} = 1 + \alpha \chi_Q
\end{equation}

where $m_0$ is the free electron mass. $\alpha$ contains various pre-factors and is arbitrarily set to 1. We can thus model the qualitative effect of mass-enhancement on the orbital-limiting field using:

\begin{equation}
\begin{split}
&H_{\text{orb}}(A=52\ \textrm{mK},T) =\\
&H_{\text{orb}}(0,T) \left( \frac{1+\alpha \chi_Q(A=52\ \textrm{mK})}{ 1+\alpha \chi_Q(A=0\ \textrm{mK})} \right)
\end{split}
\end{equation}
The quadrupolar susceptibility $\chi_Q$ increases as temperature is decreased, see inset of Extended Data Fig.~\ref{fig_SE:alternative_Hc2_mech}. This would enhance $H_{\text{orb}}$, and thus cannot account for the observed suppression of superconductivity.

Enhanced spin susceptibilities in the vicinity of magnetic phase transitions can stabilize superconductivity~\cite{Monthoux2007}. We examine whether the enhanced quadrupolar susceptibility $\chi_{Q}$ can similarly affect superconductivity in \ce{PrOs4Sb12}.
As $Q_{zx}$ is the ordering quadrupole moment for $B\parallel$[100] (which is the direction of the magnetic field in our experiments), we consider $\chi_{Q_{zx}}$, given by~\cite{Zwicknagl2009}:
\begin{equation}
\begin{split}
&\chi_{Q_{zx}} = \\
&\sum_{m=0}^{m=6} \sum_{n=6}^{n=24} \frac{2 \lvert \exp{(-E_{m}/T)} \langle \lambda_{n} \rvert \hat{O}_{zx} \lvert \lambda_{m} \rangle \rvert ^2 } {E_n - E_m} \\
&\times \sum_{m=0}^{m=6} \frac{1}{\exp{ (-E_{m}/T)}}
\end{split}
\label{quad_susc}
\end{equation}
in which $E_i$ is the mean-field energy of the hyperfine levels, and the sums are over all six hyperfine ground states $m$ and all 18 hyperfine excited states $n$.
Following Monthoux \textit{et al.}, this susceptibility could set the size of the superconducting pairing interaction strength~\cite{Monthoux2007}. We assume $\chi_Q$ to be proportional to $H_{c2}$, and simulate the effect of hyperfine interactions on $H_{c2}$ through this quadrupole susceptibility using:
\begin{equation}
\begin{split}
&H_{\chi_{Q}}(A=52\textrm{mK},T) = \\
&H_{c2}(0,T) \left(\frac{\chi_Q(A=52\ \textrm{mK})}{ \chi_Q(A=0\ \textrm{mK})}\right)
\end{split}
\end{equation}
As hyperfine interactions cause $\chi_{Q_{zx}}$ to increase with decreasing temperature (see inset of Extended Data Fig.~\ref{fig_SE:alternative_Hc2_mech}), we expect $H_{c2}$ to increase with decreasing temperature. This is in conflict with experimental results, and thus cannot explain the observed low-temperature suppression of $H_{c2}$.

We summarize the effects of hyperfine-modified Pauli and orbital limiting and quadrupolar susceptibility on $H_{c2}$ in Extended Data Fig.~\ref{fig_SE:alternative_Hc2_mech}.
\\\\
{\bf Mean-field theory}\\
To determine the mean-field behaviour of PrOs$_4$Sb$_{12}$, we have closely followed the work by Shiina~\cite{Shiina2004}, Shiina and Aoki~\cite{ShiinaAoki2004}, Kohgi~\textit{et al.}~\cite{Kohgi2003} and Kusunose~\textit{et al.}~\cite{Kusunose2009}.
In agreement with their work, dipole-dipole interactions are not included in our model because the (crystal field) excitons in \ce{PrOs4Sb12} are quadrupolar in character~\cite{Kuwahara2005, Kaneko2007} and including significant spin-spin interactions does not give an accurate description of the behaviour of the system
~\cite{Kaneko2007}.
Similar to McCollam~\textit{et al}.~\cite{McCollam2013}, we have added nuclear interactions to the mean-field Hamiltonian and extended the basis to include the Pr nuclear spin.

Our mean-field Hamiltonian consists of the crystal electric field (CEF) Hamiltonian, mean-field quadrupole-quadrupole interaction, electronic Zeeman, and nuclear interactions. The terms in the nuclear Hamiltonians are given below:

\begin{equation}\label{eqn:mf}
\begin{split}
\ham_{\textrm{HF}} &= A \hat{\vec{I}} \cdot \hat{\vec{J}} = A \left( \hat{I}_{z} \hat{J}_{z} + \frac{1}{2}(\hat{I}_{+} \hat{J}_{-} + \hat{I}_{-} \hat{J}_{+} )\right) \\
\ham_{\textrm{NQI}} & = P \hat{I}_z^2 - \frac{P}{3}\hat{I}^2 \\
\ham_{\textrm{Z, nuclear}} & = \frac{g_N \mu_N}{k_B} \hat{\vec{I}} \cdot \vec{B}
\end{split}
\end{equation}
in which $\ham_{\textrm{HF}}$ is the hyperfine interaction with hyperfine constant $A=52$~mK~\cite{Kondo1961, Bleaney1963}; $\ham_{\textrm{NQI}}$ is the nuclear quadrupole interaction, $P$~=~7~mK~\cite{Aoki2002_gn}; and $\ham_{\textrm{Z, nuclear}}$ is the nuclear Zeeman interaction. With the Pr nuclear {\em g}-factor $g_N$~=~1.7102~\cite{Macfarlane1982}, the prefactor amounts to only $(g_N \mu_N)/k_B$=0.6~mK/T. Thus, nuclear interactions in magnetic field are dominated by the hyperfine interaction.
For an overview of the basis vectors, Hamiltonians, and specific values of parameters used in our model, we refer to~\cite{Bangma_thesis}. 
In all calculations and measurements discussed in this paper, the magnetic field was oriented along the $x$-axis, $B\parallel [100]$. Experiments with $B\parallel[110]$ also show the relevance of hyperfine effects~\cite{McCollam2013}. Indeed, calculations with magnetic field along [110] or [001] yield qualitatively similar results to those described here, but the precise low-temperature shape of the \ac{afq} phase boundary changes slightly for different field directions. This is because the composition of the energy levels depends on magnetic field orientation, due to absence of four-fold symmetry on the Pr atom in \ce{PrOs4Sb12}.

The mean-field quadrupole-quadrupole interaction constant was set to 240~mK, to give a match between the lower \ac{afq} phase boundaries from our experimental data and the mean-field model (including hyperfine and nuclear terms) at 100~mK (see Fig.~\ref{fig1}(c)). However, a match at 100 mK leads to disagreement between the experimental and mean-field phase boundaries at other temperatures. Most significantly, our mean-field model overestimates the maximum critical temperature of the \ac{afq} phase by a factor of 1.9~K/1.2~K. A similar discrepancy is reported by other authors for \ce{PrOs4Sb12}~\cite{Kohgi2003, Shiina2004, Kusunose2009}, and is also observed in mean-field models of other materials~\cite{Tallon2011, Dollberg2022}. We were unable to find a value of the quadrupole-quadrupole interaction strength that resulted in a good match of experimental and mean-field phase boundaries at all temperatures: a weaker (stronger) quadrupole-quadrupole interaction reduces (increases) the maximum critical temperature, but results in a strong mismatch of the low-(high-)temperature phase boundaries. 
This mismatch is likely to be because the high temperature AFQ phase boundary is strongly suppressed by fluctuations, which are not taken into account in the mean-field theory. 

\newpage
\onecolumngrid

\section*{Extended data}

\clearpage

%

\end{document}